**„Investigations about Science Misconceptions"**

M. Risch

**Abstract**

Students entering engineering in universities of applied sciences were tested to see whether they understood easy Science problems within the first week of the first semester. The test comprised of mathematics and physics questions in both linear and diagrammatic form. Science misconceptions are investigated here by comparison to a model of misconception based on the findings of other investigators. The misconceptions demonstrated by the answers confirm this model of misconception. The results are discussed with traditional theories about education as well as with recent psychological research and with neuroscience insights. These comparisons enable further investigation of misconceptions.

**KEY WORDS**

| | |
|---|---|
| conceptual structure | problem representation |
| knowledge process | procedural knowledge |
| misconceptions | schemata |
| pre-existing knowledge | Science teaching |

**Introduction**

The education and cognitive sciences have given intensive consideration to naive beliefs as well as misconceptions in both physics (Reif 1986; Bao 2002) and mathematics (Briars 1983; Culotta 1992).

Misconceptions about everyday life phenomena are common, for example the misconception that water waves piling up on the shore, which is misinterpreted. Observation from a suitable standpoint (an airplane or a cliff) would demonstrate that the wave velocity and thus the wavelength decreases while the wave enters shallow waters piling up these waves, such an observation is difficult to make and is avoided. Thus a misconception evolves, the water waves piling up at the shore is put in compartments like „the waves coming up from below the sea". This example demonstrates how misconceptions evolve during the explanation of simple problems by overlooking adjacent observations and by neglecting critical thought.

This has been investigated a lot. Misconceptions in the minds of students about the movement of objects were studied in depth by McCloskey (1980, 1983) Halloun et al. (1985, 1987) and Rebello (2004). Misconceptions in mathematics have been studied by Reif (1987), and Resnick (1985), in physics other than mechanics by Goldberg (1987), Licht (1990), McDermott (1991), and Saxena (1992).

Psychological viewpoints of students´ difficulties in the sciences have been considered, such as concepts (Posner 1982; Reif 1986, 1987; Driver 1989), schemata (Chi 1981; Mestre 1991), representations (Wilkening 1991; Reif 1995; Lorenzo 2005), procedural knowledge (van Heuvelen 1991), cognitive anchors (Clement 1989; Laws 1997; Hammer 2000), scripts (Larkin 1980; Caramazza 1981), and curriculum (Hammer 1987; 2000; McDermott 1992).

Large surveys about overcoming students` misconceptions by new teaching methods have been undertaken, such as the IUPP (Ridgen 1993; diStefano 1996(a); diStefano 1996(b); Coleman 1998 and Hestenes 1998), the CASE study (van Heuvelen 1991) and the OCS study (Gautreau and Novemsky 1997).

The purpose of this study is further investigation on misconceptions by comparing answers to questions in either linear or diagrammatic form or with different figures about the same misconception. Research into students´ difficulties in the sciences is evaluated here. An attempt has been made to confirm well-known misconceptions as well as cognitive anchors and find new ones. In evaluating the results, first the answers to all the questions will be commented upon and then correlation of the answers to each other will be investigated. Science misconceptions will be investigated here by a four-step model of misconception derived from literature and confirmed in the answers to the questions. The results are discussed in relation to traditional theories about education as well as recent psychological and neuroscience insights about learning and memory (Cahill 1994, Erk 2003). These comparisons are useful for further investigation of misconceptions.

**The study**

941 students at universities of applied sciences were given a questionnaire comprising four pages and were asked to answer anonymously and spontaneously by marking appropriate boxes. Students were told that neither calculations nor sophisticated knowledge was necessary to answer. This questionnaire was to be answered by the students within 15 minutes; it consisted of 4 sections corresponding to 4 pages as follows: six questions on school education, three questions on mathematics (abstraction, algebra, analysis), two questions on mechanics and three questions on optics, acoustics, and electricity.

**Methods**

Students participating in the study were engaged in different kinds of engineering, combined economics and engineering, or design and engineering courses at four German universities of applied sciences (Amberg-Weiden, Augsburg, Munich, and Nuremberg). All 941 students were at the very beginning of their studies in the first week after terms began. For comparison, eight education students at the end of their courses before the start of their teacher training in primary and secondary schools were given the questionnaire. They are referred to here as "teachers".

**Results of Mathematics Questions**

The first question in the mathematics section was to test the cognitive mathematical abilities independent of school mathematics instruction. The question about exponential growth of a paper to be folded is described in a standard psychology textbook (Zimbardo 1988). Students restricted to visual imagery (Sternberg 1985; diSessa 1993) will overlook exponential growth and guess a relatively small thickness of the folded paper. Application of mathematical symbols or concepts will overcome visual blockbustering (Adams 1979; diSessa 1993) and lead to a result close to reality (millions of kilometers).

Question 1 *(figure 1)*:
**Fig. 1:**

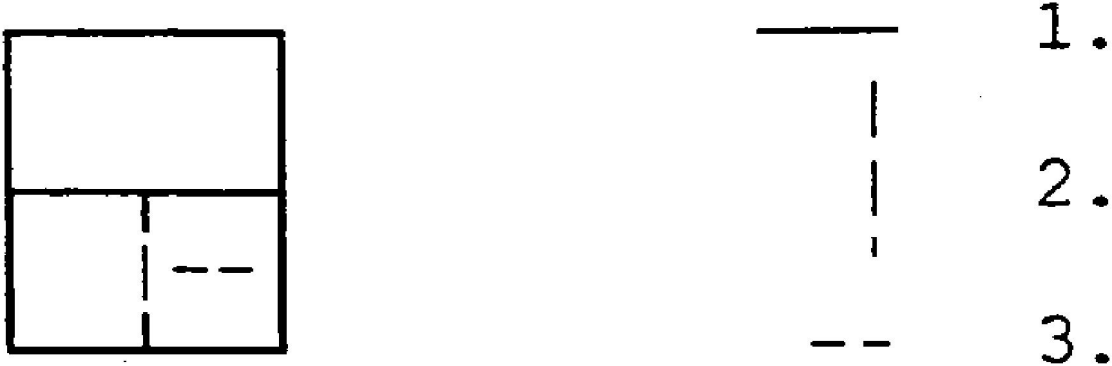

**Fig. 1. Question 1:** If this paper here is folded 40 times each time in the middle, how thick will it become?

**Table I  Answers to abstraction question**

(Correct answer marked bold)

| Answer | students | % | teachers | % |
|---|---|---|---|---|
| failed to answer | 9 | 1.0 | – | – |
| will become some mm thick | 101 | 10.7 | 1 | 12 |
| will become some cm thick | 446 | 47.4 | 2 | 25 |
| will become some m thick | 167 | 17.8 | 1 | 12 |
| will become some km thick | 90 | 9.6 | 1 | 12 |
| **will be thousands of km thick** | **114** | **12.1** | **3** | **38** |
| I cannot conceptualize the question | 14 | 1.5 | – | – |

These answers show a rather naive flat-minded way of thinking avoiding the complications of grasping the exponential. Here imagination is required rather than observation, the misconception is caused by shifting imagination from difficult exponential thinking to straightforward linear thinking.

The second question in the mathematics section was to test Analysis abilities. Three vectors forming a regular triangle (Reif 1987) were marked by arrows in both text and picture, and a student with Analysis understanding should recognize the angle between two vectors as the angle between directions, 120°, as indicated by arrows (Jagannathan 1997). However, naive beliefs or visual imagery (Zimbardo 1988) will make a student simply see a triangle with a 60° angle.

Question 2: What is the approximate angle between vectors A and B? *(A drawing of a triangle formed by vectors carrying arrows was provided)*

**Table II  Answers to mathematics Analysis question**

| Angle | students | % | teachers | % |
|---|---|---|---|---|
| failed to answer | 3 | 0.3 | – | – |
| 0° | 1 | 0.1 | – | – |
| 60° | 781 | 83.0 | 5 | 62 |
| 90° | 8 | 0.8 | – | – |
| **120°** | **104** | **11.0** | **3** | **38** |
| 180° | 41 | 4.4 | – | – |
| I cannot conceptualize question | 3 | 0.3 | – | – |

Here again imagination is required rather than observation, the misconception is caused by shifting imagination from difficult vector space thinking to straightforward thinking. The last question in the mathematics section was to test algebra abilities. The question concerned a graph of an arbitrary function with four marked sections A to D, A with a point without derivative (sharp turn), B with saddle, C with relative maximum and D with a zero. Students were asked to determine which sections have a point without slope (= derivative becoming zero), which requires that the student understands the use of infinitesimal in the formation of a derivative.

Question 3 *(figure 2)*:

**Fig. 2.**

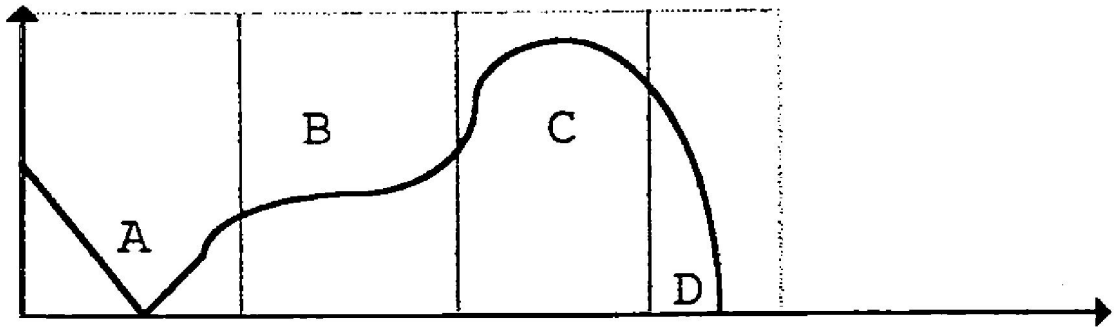

**Fig. 2. Question 3:** A mathematical function is plotted in the graph below. In which of the sections is the slope of shown curve zero (mark only one)?

**Table III   Answers to mathematics algebra question**

| Answer | students | % | teachers | % |
|---|---|---|---|---|
| failed to answer | 12 | 1,3 | - | - |
| A and B | 50 | 5,3 | - | - |
| **B and C** | **384** | **40,8** | **7** | **88** |
| A and C | 162 | 17,2 | - | - |
| A and D | 78 | 8,3 | - | - |
| A, B and C | 193 | 20,5 | 1 | 12 |
| B, C and D | 16 | 1,7 | - | - |
| A, B, C, and D | 31 | 3,3 | - | - |
| I cannot conceptualize „slope" | 15 | 1,6 | - | - |

Many students confused the zero value of a function with the zero value of derivative by applying a false thinking shortcut as described by cognitive psychology (Reif 1987; Hammer 1996). Here again imagination is required in addition to observation, the misconception is caused by shifting imagination from difficult infinitesimal thinking to straightforward thinking in lines, points and curves.

**Results of Mechanics Questions**

The first mechanics question concerned the movement of a rocket coasting in space as described by Halloun and Hestenes (1985), Gunstone (1987), Laws (1991) as well as Rebello (2004). The questions were to test the students´ understanding of Newtonian principles of mechanics, as opposed to naive beliefs. Such preconceptions believe an „impetus" kind of force was necessary to maintain constant motion, thus violating the Galilean principle of inertia. To differentiate students´ difficulties this question was subdivided into three sub- questions.

Question 4a *(figure 3)*:

**Fig. 3.**

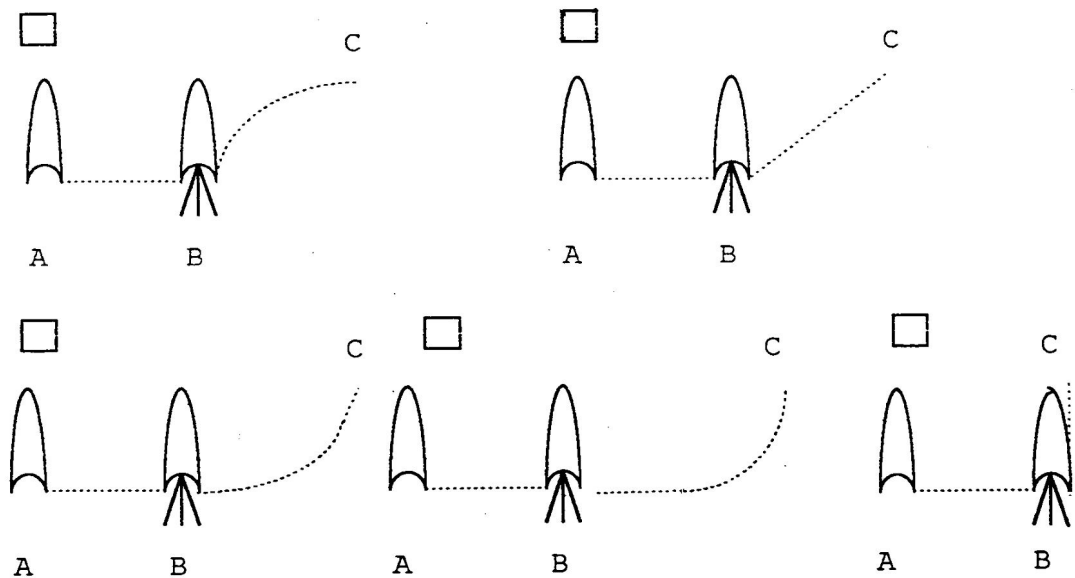

**Fig. 3. Question 4a:** A rocket is coasting in space without any forces from A to B, in B engines are ignited to full thrust perpendicular to coasting motion, which of the ways shown will the rocket follow from B to C:

**Table IV  Answers to mechanics rocket question part A**

| Answer | students | % | teachers | % |
|---|---|---|---|---|
| failed to answer | 8 | 0.9 | – | – |
| (sharp turn at ignition) | 56 | 6.0 | 1 | 12 |
| (straight line) | 179 | 19.0 | 1 | 12 |
| **(curved upward without turn)** | **519** | **55.2** | **6** | **75** |
| (curved upward, first straight) | 92 | 9.8 | – | – |
| (perpendicular upward) | 86 | 9.1 | – | – |

Rather than following the principles of Newtonian mechanics (forces causing accelerated motion, curved upward in this case), many students seem to rely on naive „impetus" theory with a constant straight line motion caused by a force as taught by medieval physicists (Ibn Sina 1885; Abu´l Barakat 1939). Also, some students seem to believe in a retarded motion (force first, motion later, curved upward after a straight line in this case) as is frequently shown in cartoons, this „warming up"-misconception is discussed by diSessa (1993, p. 133; 1998). These misconceptions contradict Newtonian laws and seem not to be eliminated by thorough school instruction, because of the strength and ease of visual imagery as opposed to the task of applying scientific concepts (Hewitt 1983; McDermott 1991). The answers reflect a simplified way of thinking („p-prim" diSessa 1993, 1998) avoiding the complications of grasping the acceleration as a differential of motion. To work out these difficulties, the rocket question was elaborated in a second part:

Question 4b: While the rocket is moving from B to C, what happens to speed?

**Table V  Answers to mechanics rocket question part B**

| Answer | students | % | teachers | % |
|---|---|---|---|---|
| failed to answer | 8 | 0.9 | – | – |
| speed is constant | 151 | 16.0 | 1 | 12 |
| **increases steadily** | **451** | **47.9** | **5** | **62** |
| decreases steadily | 2 | 0.2 | – | – |
| increases first, then constant | 319 | 33.9 | 2 | 25 |
| constant first, then decreases | 4 | 0.4 | – | – |
| I cannot conceptualize quest. | 5 | 0.5 | – | – |

In this part of the rocket question, asked in words rather than pictograms visualizing the problem as in the preceding part, even more students fell into the impetus thinking (constant force causes constant motion), confirming theories of „prevailing misconceptions" (McDermott 1984). To further differentiate between word and pictogram questions, a third part of the rocket problem was designed with both words as well as arrows as an intermediate.

Question 4c: In C rocket engines are turned off, which way will the rocket coast thereafter? (Arrows have indicated Answers)

**Table VI  Answers to mechanics rocket question part C**

| Answer | students | % | teachers | % |
|---|---|---|---|---|
| failed to answer | 11 | 1.2 | – | – |
| straight in same direct. as in A | 108 | 11.5 | 1 | 12 |
| **straight in same direct. as in B** | **557** | **59.2** | **6** | **75** |
| straight upward | 123 | 13.1 | – | – |
| curved upward, bent out | 117 | 12.4 | – | – |
| curved upward, bent in | 24 | 2.6 | 1 | 12 |

In this part of the rocket question, impetus misconceptions showed up to approximately the same extent as in the first pictogram style question.

In order to investigate further into the impetus kind of naive conception about movement of objects, the last mechanics question was about a ball shot into a circular tube leaving the end of the tube without the influence of gravity (McCloskey 1980; Mestre 1991; and van Heuvelen 1991 a). In order to clarify the influence of visual imagery, the question was repeated with three half turns rather than one half. Tubes with one and three half turns were shown as pictures (figure 4), and the ball shooting as arrows.

Question 5 *(figure 4)*:
**Fig. 4.**

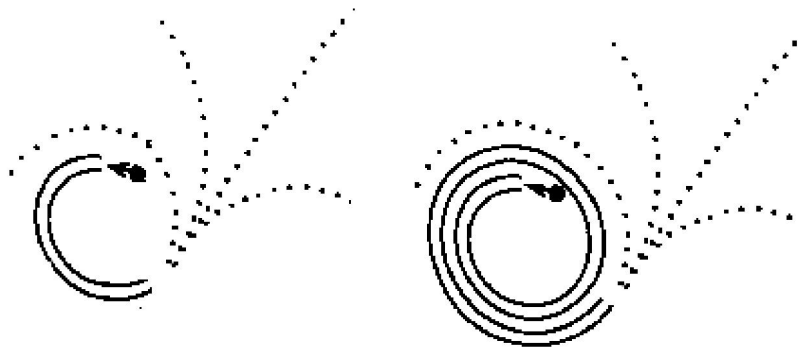

**Question 5:** Balls are shot at a high speed into each of the circular tubes as shown by arrows leaving the tube at the other end with same speed. With no external forces such as gravity or friction, indicate the further path of balls for each tube.

**Table VII  Answers to mechanics circular motion question**

| Answer | students | % | teachers | % |
|---|---|---|---|---|
| failed to answer | 29 | 3.1 | - | - |
| **straight both times** | **713** | **75.8** | **5** | **62** |
| straight with half turn, bent up with three half turn tube | 97 | 10.3 | 1 | 12 |
| bent upward with half turn tube straight with three half turn t. | 3 | 0.3 | 1 | 12 |
| curved upward both tubes | 24 | 2.6 | 1 | 12 |
| curved downward, one or both turn | 74 | 7.9 | - | - |

These answers clearly prove the importance of visual imagery; since the image of a three half-turn tube induces students more strongly towards an impetus misconception of movement than an image of a one half turn tube. The more turns seen by the students, the stronger the false belief in an ever-lasting circular motion without cause as described by Aristotle. The answers reflect a naive way of thinking avoiding the complications of grasping the acceleration as a cause of circular motion. An entire 8% of students did not read the question carefully and included effects of gravity, though gravity was definitely excluded in the question.

**Results of Physics Questions**

The first of three physics questions other than mechanics was an optical question of a picture formed by a single convex lens described by Goldberg and McDermott (1987). This question was designed to test the concept that intersecting rays forms optical images. The questionnaire handed to the students contained a picture with a bulb with U-type glowing wire, a converging lens and a screen with the image of a glowing wire turned upside down.

Question 6a *(figure 5)*:

**Figure 5.**

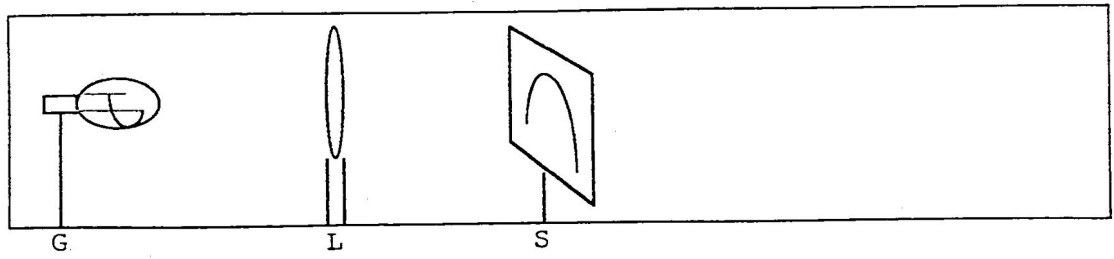

glowing wire in bulb    lens    screen with image of glowing wire

**Question 6a:** What happens when the lens is removed?

**Table VIII   Answers to optics (lens) question A**

| Answer | students | % | teachers | % |
|---|---|---|---|---|
| failed to answer | 47 | 5.0 | - | - |
| image upside, not upside down | 289 | 30.7 | 4 | 50 |
| **image vanishes** | **323** | **34.3** | **4** | **50** |
| image does not change | 6 | 0.6 | - | - |
| image becomes less sharp | 252 | 26.8 | - | - |
| I cannot conceptualize „lens" | 23 | 2.4 | - | - |

In spite of ray optics training at school, many students are not aware that the lens forms the image. The few students not trained marked "cannot conceptualize lens". Students rather adhere to misconceptions such as the lens turns image upside down or sharpens image. The answers show a simple way of thinking avoiding the complications of the concept of inter-secting rays causing picture forming. In order to investigate these misconceptions further, the question was repeated in an altered form:

Question 6b: What shows up when half of the lens is covered?

**Table IX  Answers to optics (lens) question B**

| Answer | students | % | teachers | % |
|---|---|---|---|---|
| failed to answer | 59 | 6.3 | - | - |
| image vanishes | 47 | 5.0 | - | - |
| image becomes less sharp | 72 | 7.7 | 1 | 12 |
| image becomes half | 618 | 65.7 | 2 | 25 |
| **image does not change** | **14** | **1.5** | **-** | **-** |
| image upside, not upside down | 9 | 1.0 | - | - |
| **image becomes darker** | **121** | **12.9** | **5** | **62** |

More than in the first part of the question, misconceptions about ray optics were evident. Students seem to imagine a lens visually as a kind of valve letting light through, rather than deflecting rays, which combine to form an image. In order to investigate these misconceptions further, another question about image forming was included:

Question 6c: What happens when the screen is moved towards the lens?

**Table X  Answers to optics (lens) question C**

| Answer | students | % | teachers | % |
|---|---|---|---|---|
| failed to answer | 66 | 7.0 | - | - |
| image vanishes | 24 | 2.6 | - | - |
| **image becomes less sharp** | **422** | **44.8** | **7** | **88** |
| image becomes bigger | 346 | 36.8 | - | - |
| image becomes darker | 3 | 0.3 | - | - |
| image does not change | 15 | 1.6 | - | - |
| image upside, not upside down | 64 | 6.8 | 1 | 12 |

Here students seem to overlook that a sharp image will form only in one or two locations and that the magnification de-pends on the position of lens, not the screen.

The second question was a question about the sound of a resonating string, the length of which will be doubled. A pictogram of string and doubled string was shown and making an analogy with a guitar helped students.

Question 7: When the length of string is doubled, what happens to the sound (pitch):

**Table XI  Answers to acoustic (string) question**

| Answer | students | % | teachers | % |
|---|---|---|---|---|
| failed to answer | 28 | 3.0 | - | - |
| sound becomes higher (frequency) | 65 | 6.9 | 2 | 25 |
| **sound becomes lower (frequency)** | **719** | **76.4** | **6** | **75** |
| sound becomes louder | 23 | 2.4 | - | - |
| sound becomes weaker | 6 | 0.6 | - | - |
| higher and louder | 2 | 0.2 | - | - |
| higher and weaker | 2 | 0.2 | - | - |
| lower and louder | 36 | .3.8 | - | - |
| lower and weaker | 59 | 6.3 | - | - |

The answers indicate that understanding a vibrating string is a cognitive anchor with easy visualization and possibility for application of readily available pre-existing knowledge (Clement 1989). However, more than 10% of students marked two boxes although they were asked to mark just one for a question.

The last question was about a simple electric DC circuit of five lamps and a battery similarly described by Hewitt (1983), Licht (1990), McDermott and Shaffer (1992), Engelhardt (2004), and Cepni (2006). A simple circuit diagram showing a single lamp, two lamps in parallel and two lamps in a series all con-nected in parallel to a battery were shown in the questionnaire. In order to investigate student's comprehension of current and voltage, they were asked if these identical lamps shine equally brightly or more or less brightly or not at all. Question 8a-d *(figure 6)*:

**Fig. 6.**

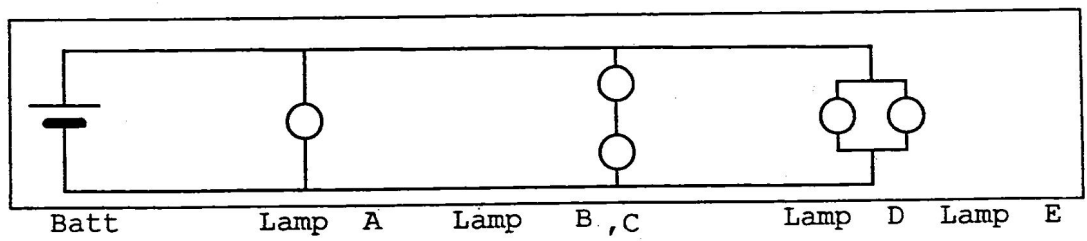

**Fig. 6. Question 8a-d:** A battery is connected to 5 equal lamps as shown. These identical lamps can shine by flow of current either brightly, less brightly or not at all.

**Table XII  Answers to electricity questions** (Question 8a, b, c, and d)

| Lamp A (single) will shine: | students | % | teachers | % |
|---|---|---|---|---|
| failed to answer | 52 | 5.5 | - | - |
| **brightly** | **836** | **88.8** | **8** | **100** |
| less brightly | 48 | 5.1 | - | - |
| not at all | 4 | 0.4 | - | - |
| Lamp B (series with C) will shine | | | | |
| failed to answer | 55 | 5.8 | - | - |
| brightly | 222 | 23.6 | 1 | 12 |
| **less brightly** | **598** | **63.6** | **7** | **88** |
| not at all | 65 | 6.9 | - | - |
| Lamp C (series with B) will shine | | | | |
| failed to answer | 66 | 7.0 | - | - |
| brightly | 184 | 19.6 | 1 | 12 |
| **less brightly** | **620** | **65.9** | **7** | **88** |
| not at all | 70 | 7.4 | - | - |
| Lamps D and E (in parallel with each other) will shine | | | | |
| failed to answer | 59 | 6.3 | - | - |
| **brightly** | **563** | **59.8** | **6** | **75** |
| less brightly | 250 | 26.6 | 2 | 25 |
| not at all | 68 | 7.2 | - | - |

Students are subject to pitfalls, such as the idea that the current is used up upon transport through wires or those two lamps in parallel divide in the current which one identical lamp would use. Similar naive conceptions have been found by Mestre (1991), McDermott (1991, 1992), Saxena (1992), Millar (1993) Reif (1995), Engelhardt (2004), and Cepni (2006). These answers show a way of thinking avoiding the complications of applying the concepts of current and voltage.

**Discussion: Correlation of Results**

The average percentage of correct answers was measured in each of the three parts. In the mathematics part, a point was granted for each of the two questions if correct, in the physics part for each of the twelve questions. In the abstraction part, up to 4 points have been awarded for the answer for statistical reasons (4 for thousands of kilometers, 3 for kilometers, 2 for meters, 1 for cm and 0 for mm or no answer). Maximum is four for abstraction, two for mathematics and 12 for physics. Correlation between average percentages of correct answers (e.g. between abstraction and mathematics) was calculated by statistical methods (standard deviation and correlation coefficients).

**Table XIII Results and correlation:** average percentage of correct answers

| University | engineering students | Education students |
|---|---|---|
| number of students | 941 | 8 |
| Performance in Mathematics | 30.0% correct | 62.5 |
| Performance in Abstraction | 40.2% of max. points | 59.5 |
| Performance in Physics | 57.2% correct | 75.0 |
| Standard deviation Mathematics | 29% | 35% |
| Standard deviation Abstraction | 29% | 40% |
| Standard deviation Physics | 20% | 15% |
| Correlation Coefficient Mathematics to Abstraction | 0.14 | 0.8 |
| Correlation Coefficient Abstraction to Physics | 0.16 | 0.2 |
| Correlation Coefficient Physics to Mathematics | 0.18 | 0.2 |

The education students at the end of their studies performed considerably better in both mathematics and physics, however, because of the small number the statistical conclusion is not very significant.

**Conclusions**

This research confirms misconceptions in mathematics and physics as well as the existence of cognitive anchors. A new misconception in mathematics and a new cognitive anchor in physics have been found. The concept change, which is necessary to overcome misconceptions, is difficult for students because it requires a change from the ontological categories matter or things to processes and mental states (Chi 1994). For example, question 6, students seem to imagine a lens as a kind of valve letting light through, rather than deflecting rays which combine to form an image. Central concepts are likely to be rejected when they have generated a class of problems which they appear to lack the capability to solve (Posner 1982), such as Newtonian mechanics which is perceived and represented by students to be in the presence of frictional forces, questions 4 and 5, and thus seem to contradict everyday life experiences with overwhelming influence of frictional forces. Representations of problems in students may be in the form of propositions or images (Posner 1982), which can prevent application of central concepts, such as in questions 1 to 3 and 5. Science misconceptions can be explained by comparison to a simple four step cognitive model of misconception found in children as well as in adults. This model was developed with application of Piaget´s ideas about assimilation and reconciliation when encountering new phenomena in nature (Posner 1982; Chi 1994). In the development of knowledge, intellectual norms have to be used according to Piaget´s epistemology (Palmer 2007; Piaget 2000), such as autonomy, entailment, inter-subjectivity, objectivity, universality (mnemonic AEIOU).

- autonomy- use of own reasoning
- entailment (necessary knowledge) - a necessary relation about what has to be
- inter-subjectivity - being in line with generally accepted axioms which are a paradigm case of common ground between different thinkers
- objectivity - being justified as a true response in a valid

argument

- universality - whether or not open to transfer under different causal conditions

These required norms for reasoning imply pitfalls, however. Though autonomy is a condition for reasoning it can evoke wrong representations following naive conclusions drawn from observations, which lead to misconceptions. According to Piaget, children have a tendency to adapt new observations to old naïve beliefs and misconceptions (called assimilation by Piaget) rather than having a conceptual change to new concepts explaining the phenomenon better (called accommodation by Piaget). This behavior is explained by the tendency of students to reconcile new observations with old misconceptions (Posner 1982; diSessa 1993). The misconceptions discussed can be analyzed by a model of four cognitive steps, which are based on Posners´ ideas (1982), a mnemonic is RACA:

- ■ 1. Rejection (R): Rejection of observational theory, (example: the individual observes water waves piling up at shore, since the height of waves seems to be much less on the ocean far away, the magnitude of waves seems to come from „elsewhere", the ocean depth. This source is not easily observable and therefore it is rejected)

- ■ 2. Avoid concern (A): lack of concern with experimental findings, (example: careful observation from a suitable standpoint, an airplane or a cliff, would yield to correct information, the wave velocity and thus the wavelength decreases while the wave enters shallow waters, such an observation is difficult and is avoided)

- 3. Compartmentalization (C): A compartmentalization of knowledge to prevent information from conflicting with existing belief, (example: the false explanation could be tested by careful observation, however, as it is difficult to make, this investigation is neglected and a misconception evolves, the interpretation of water waves piling up at shore which is put in compartments like „the waves coming up from below the sea").
- 4. Assimilation (A): Assimilation of new information into existing naïve concepts, (example: in absence of objections or critical thought the false explanation will become a false representation)

All the observed misconceptions of students can be explained by these four cognitive features of recognition.

The misconceptions and poor comprehension in mathematics and sciences found in first year students lead to the conjecture that more emphasis should be on concepts like those outlined by Reif (1987), Purcell (1997) and Griffiths (1997). This conclusion is strengthened by the connection found between misconception and visualization error avoiding the change from the ontological categories matter or things to processes and mental states (Chi 1994). Since misconceptions are seemingly founded on false preconceptions, more emphasis in Science teaching should be on hands-on experiments, blackboard drawings without formulas and concept oriented teaching in sciences. This might ease the problems found, as has been concluded by diStefano (1996 a, b) and Gautreau (1997). One sin-gle experiment can change the situation and induce concept change (Abott 2000; Bao 2004). Involving emotional events in the process of teaching improves memory and eases overcoming misconceptions (Cahill 1994). Therefore suspense stories like "How Galileo overcame superstitious believes" raise emotions improving memory and attendance thus easing concept change, according to results of psychological as well as neurological research (Erk 2003). Emotional events improve memory but they seem not to prevent misconceptions. When the rocket question (number 4 A to C) was answered by students the same misconceptions showed up when the question was described by a picture and explained with words which can raise emotions as when the question was put in a linear form.

The four elements of the misconception model were confirmed by the answers which show a simple way of thinking avoiding complications of grasping the definitions of physics and mathematics, avoiding the required difficult change from the ontological categories matter or things to processes and mental states (Chi 1994), according to the model.